\documentclass[twocolumn,prb,aps,amsmath,amssymb,superscriptaddress,showpacs]{revtex4-1}
\usepackage{dcolumn}
\usepackage{hyperref}
\usepackage{multirow}
\usepackage{booktabs}
\usepackage{graphicx}
\usepackage{tabularx}
\hypersetup{
    colorlinks=true,
    linkcolor=blue,
    filecolor=magenta,      
    urlcolor=black,
    citecolor=blue,
}
\makeatletter
\g@addto@macro\normalsize{%
  \setlength\abovedisplayskip{12pt}
  \setlength\belowdisplayskip{12pt}
  \setlength\abovedisplayshortskip{12pt}
  \setlength\belowdisplayshortskip{12pt}
}
\makeatother
\usepackage{physics}
\usepackage{epsfig}
\usepackage{graphics}
\usepackage{float}
\usepackage{amsmath}
\usepackage{xcolor,colortbl}
\usepackage[utf8]{inputenc}
\usepackage[T1]{fontenc}
\usepackage{lmodern} 
\usepackage{soul}
\usepackage[version=4]{mhchem}
\usepackage{color}
\usepackage{soul}

\definecolor{Gray}{rgb}{0.72,0.72,0.98}
\definecolor{LightCyan1}{rgb}{0.83,0.83,0.98}
\definecolor{LightCyan2}{rgb}{0.91,0.92,1}

\begin{document}

\title{Tunable Interlayer Excitons in Bilayer Graphene Nanoribbons}

\author{Alexandre R. Rocha}
\affiliation{
	Instituto de Física Teórica, Universidade Estadual Paulista (UNESP),\\
	R.\ Dr.\ Bento Teobaldo Ferraz, 271, São Paulo, 01140-070 São Paulo, Brazil.}
\author{Rodrigo G. Amorim}
\affiliation{Departamento de F\'{i}sica, ICEx, Universidade Federal Fluminense - UFF, Volta Redonda/RJ, Brazil}

\author{Wanderl\~{a} L. Scopel}
\affiliation{Departamento de F\'{i}sica, Universidade Federal do Esp\'{i}rito Santo- UFES , Vit\'{o}ria/ES, Brazil}

\author{Cesar E. P. Villegas}
\affiliation{ Departamento de Ciencias, Universidad Privada del Norte, Lima 15434, Peru}

\date{\today}

\begin{abstract}
	Vertically stacked van der Waals structures are promising platforms that enable layer engineering, opening new avenues for the quantum control of elementary excitations, including optically generated bound electron-hole pairs. Here we employ excited-state density functional calculations to demonstrate strong interlayer excitonic coupling in one-dimensional van der Waals nanostructures derived from armchair graphene nanoribbons.
	The excitonic response exhibits prominent peaks in the near-infrared range, mainly attributed to intralayer excitons, while interlayer excitations with absorption peak strengths of up to 13\% of the maximum absorption are also observed. Both type-I and type-II band alignments are found, which promote the formation of intralayer and interlayer excitons. Notably, interlayer excitons in these systems exhibit long-lived radiative lifetimes at room temperature, ranging from 1 nanosecond to 9.4 microseconds. Our calculations suggest the potential to tune the excitonic response and lifetimes of bilayer graphene nanoribbons via careful engineering of the stacking order.
\end{abstract}

\maketitle


\section{Introduction}
Atomically thin semiconductors have emerged as promising candidates for the next generation of flexible and wearable electronic and optoelectronic devices \cite{katiyar20232d,moses20212d,cheng20212d,shahbaz2024advancements}. Due to the reduced dielectric screening of the surrounding medium, these systems exhibit enhanced Coulomb interactions \cite{cudazzo2011dielectric,latini2015excitons,villegas2024screened}, which interact with other degrees of freedom, such as spin, valley, or layer indices. This can give rise to fascinating excitonic effects that dominate their optical response \cite{yu20172d,xiao2017excitons,schaibley2016valleytronics,wang2018colloquium}. 

By vertically stacking a second two-dimensional (2D) semiconductor, interlayer excitons—characterized by electrons and holes separated across the two layers—can arise \cite{wilson2021excitons,li2022interlayer,ovesen2019interlayer}. These excitons, formed either through electrical charge transfer via type-II band alignments or interlayer hybridization of band-edge states, have been observed in both heterobilayer and homobilayer  systems of transition metal dichalcogenides (TMDCs)  \cite{jiang2021interlayer,gerber2019interlayer,deilmann2018interlayer}.
In general, interlayer excitons formed in heterobilayers exhibit excitation energies lower than those of their individual monolayers, very small oscillator strengths, and exceptionally long lifetimes, ranging from several tens to hundreds of nanoseconds \cite{miller2017long,nagler2017giant}. In contrast, interlayer excitons in homobilayer systems are highly tunable by out-of-plane electric fields \cite{wang2018electrical,peimyoo2021electrical}. These excitons also have transition energies that lie above the intralayer one \cite{deilmann2018interlayer,gerber2019interlayer}, exhibit relatively large oscillator strengths—up to 20\% of the intralayer one \cite{gerber2019interlayer}—and lifetimes of the order of tens of nanoseconds \cite{palummo2015exciton,wang2018electrical}. These features make interlayer excitons in homobilayer systems optically bright and enable their use for several light-detection applications.

Interestingly, interlayer excitons have also been experimentally observed in 3D bulk van der Waals materials, exhibiting oscillator strength intensities and excitation energies similar to those of their bilayer counterparts \cite{arora2017interlayer,arora2018valley,jindal2020interlayer,villegas2024optical}. 
However, evidence of similar effects in 1D systems is more scarce. A recent study has provided experimental evidence of long-lived intertube excitonic correlations in core/shell/skin coaxially stacked C/BN/MoS$_2$ nanotubes \cite{burdanova2022intertube}. This represents the first experimental evidence, to the best of our knowledge, for the existence of one-dimensional (1D) interlayer excitons. Given the wide range of potential applications these novel one-dimensional (1D) van der Waals systems may offer, significant attention has been directed toward understanding their fundamental electronic and excitonic properties \cite{xiang2020one,guo2021one}. 

A promising 1D van der Waals layered system for optoelectronic applications is graphene nanoribbons. Due to their relatively high mobilities, tunable band gaps, and enhanced Coulomb interactions that depend on their edges \cite{villegas2024screened,tian2023graphene,villegas2013plasmon}, these systems can offer advantages over two-dimensional systems for device applications \cite{wang2021graphene,tripathi2024optoelectronic}.
In particular,  the optical band gaps of armchair graphene nanoribbons (AGNRs) can be tuned with the number of atoms, $N$, that define their lateral width.
These AGNRs and their variations have been extensively studied both experimentally and theoretically, providing strong evidence of the key role excitonic effects have on their optical response. \cite{prezzi2008optical,tries2020experimental,denk2014exciton,villegas2014optical,villegas2024screened}. 

In light of the novelty of interlayer excitons in 2D, and the recent advances in synthesis methods for achieving precise-edge single-layer and bilayer graphene nanoribbons \cite{wang2017precision,chen2020graphene,oliveira2015synthesis},
here we perform many-body perturbation theory calculations, which accurately predicts quasiparticle properties \cite{onida2002electronic}, to study the electronic and excitonic properties of carbon-based bilayer AGNRs systems, along with their exciton radiative rates for different stacking configurations. Our results reveal that, depending on stacking, bilayer AGNRs with $\beta$-stacking are prone to exhibit both intralayer and interlayer excitons.
For bilayer systems, composed of ribbons with different widths, we report the formation of both type-I and type-II band alignments, which favor the observation of pure interlayer excitons. Notably, these excitons exhibit long-lived room-temperature radiative lifetimes ranging from 258 ns to 9.4 $\mu$s.
\section{Theory and Methodology}
Ground-state density-functional theory (DFT), as implemented in the \textsc{quantum} \textsc{espresso} package \cite{qe} was employed to calculate the structural and electronic properties of all bilayer nanoribbons systems with various widths.
The Perdew-Burke-Ernzerhof generalized-gradient approximation (GGA-PBE) and van der Waals (vdw) interactions within the Grimme-D2 scheme \cite{pbe-d2} were used to describe the exchange-correlation functional.
A kinetic energy cutoff of 90 Ry was adopted to expand the Kohn-Sham orbitals in a plane-wave basis set, while the Brillouin zone was sampled on a Monkhorst-Pack $\Gamma$-centered \emph{k}-mesh of $24 {\times} 1 {\times} 1$. 
A vacuum spacing of ${\sim}12$ \AA{} was used along the non-periodic directions to avoid spurious interactions with the periodic images of the system. Our ground-state calculations were carried out without including spin–orbit coupling effects, as we verified that their overall impact on the electronic band structure is on the order of a few tens of $\mu$eV. All structures were fully relaxed to their equilibrium positions with residual forces smaller than 1 meV/\AA{} and pressures on the lattice unit cell smaller than 0.1 kbar.

To obtain the excitonic optical response, the calculation was performed in three steps. First, plane-wave DFT calculations were carried out on the fully relaxed structures. Next, within the $G_{0}W_{0}$ approximation, the quasiparticle (QP) corrections, i.e., the electron self-energy was computed and used to correct the Kohn–Sham eigenvalues. Finally, after applying the QP corrections, excitonic effects were included by solving the Bethe–Salpeter equation (BSE). The QP energies are obtained by using the Green’s function formulation of many-body perturbation theory (MBPT), taking the first order Taylor expansion of the self-energy $\Sigma_{n\vb{k}}$ around the Kohn-Sham eigenvalues $\epsilon_{n\mathbf{k}}^{\text{KS}}$ \cite{onida2002electronic},
\begin{equation}
	E_{n\mathbf{k}}^{\text{QP}} = \epsilon_{n\mathbf{k}}^{\text{KS}} +Z_{n\vb{k}} \mel{n\mathbf{k}^{\text{KS}}}{\Sigma_{n\vb{k}}(\epsilon_{n\vb{k}}^{\text{KS}}) -V^{\text{xc}}_{n\mathbf{k}}}{n\mathbf{k}^{\text{KS}}}\,,
	\label{eq1}
\end{equation}
where $\ket{n\mathbf{k}^{\text{KS}}}$ describes the Kohn-Sham eigenstates, and $Z_{n\vb{k}}$ ($V^{\text{xc}}_{n\mathbf{k}}$) is the quasiparticle renormalization factor (DFT exchange-correlation potential). Within the $G_0W_0$ approximation, the self-energy is described as the product of the one-electron Green's function and the screened Coulomb potential, $\Sigma = iG_0W_0$. The dielectric screening matrix is constructed on  a 
\textbf{k}-grid sample of $72 {\times}1{\times} 1$, and considering the Plasmon-Pole approximation in the Godby-Needs scheme \cite{godby1989metal} with an energy cutoff of 10 Ry and including 480 bands. Accordingly, the exchange potential matrix uses a 50 Ry energy cutoff. The GW sum-over-states is performed on 600 bands, which  is sufficient to provide accurate results owing to the implementation of Bruneval-Gonze terminators \cite{bruneval2008}. 
 Due to the rectangular geometry of the systems under study, we model the Coulomb interaction using a truncated potential in a box-shaped geometry, considering a cutoff of 4 Ry to minimize spurious interactions with neighboring periodic images.

The excitonic effects on the optical spectra are analyzed by solving the Bethe-Salpeter equation (BSE) within the Tamm-Dancoff approximation, whose Hamiltonian representation for the eigenvalue problem is given by \cite{louie2000PRB}
\begin{equation}
	H^{\text{BSE}}_{\substack{vc\vb{k}\\v'c'\vb{k}'}} = \left(\epsilon^{\text{QP}}_{c\mathbf{k}}-\epsilon^{\text{QP}}_{v\mathbf{k}}\right)\var_{c,c'}\var_{v,v'}\var_{\vb{k},\vb{k}'}+ \Xi^{\text{eh}}_{vc\vb{k},v'c'\vb{k}'}\,.
	\label{eq2}
\end{equation}
Here the first term in parenthesis comprises a diagonal part that contains the
quasiparticle energy differences, and $\Xi^{\text{eh}}= K^{x} + K^{c}$ is the so-called BSE kernel, composed of an $e$-$h$ attraction term ($K^{c}$) and a repulsive exchange one ($K^{x}$). Upon diagonalizing the BSE Hamiltonian, \cite{louie2000PRB}
\begin{equation}
	\sum_{v'c'\vb{k}'}H^{\text{BSE}}_{\substack{vc\vb{k}\\v'c'\vb{k}'}}A^{S}_{v'c'\vb{k}'} = \Omega_SA^{S}_{vc\vb{k}}\,,
	\label{eq3}
\end{equation}
we obtain the exciton envelope wavefunction $A^S_{vc\mathbf{k}}$ and eigenvalues $\Omega_S$, for the $S$-\emph{th} exciton state. We have verified the validity of the Tamm-Dancoff approximation for describing the lowest-energy excitonic states in the different bilayer systems studied. The corresponding results are presented in the Supplementary Information.
For all systems, the BSE equation was solved on a fine $\vb{k}$-grid sample of $200 {\times}1{\times} 1$ points considering the six highest-occupied valence bands and the six lowest-unoccupied
conduction bands. This number of bands is sufficient for characterizing the lowest-energy excitonic states. Note however, that if one is interested in analyzing deep energy excitons, beyond the visible range, more bands should be included. The QP and BSE parameters have been carefully converged in our previous work and successfully employed to describe the band gaps and exciton binding energies of up to 12 different AGNRs belonging to different families \cite{villegas2024screened}.

By considering that  the effects of spin-orbit coupling with respect to the e-h interaction are negligible, Eq. (\ref{eq3}) can describe spin-singlet and spin-triplet excitations, the latter obtained by disregarding the exchange term in the BSE kernel \cite{wang2012quasiparticle}. 
In all calculations we consider light polarized parallel to the ribbon axis, and an artificial Lorentzian broadening of 0.04 eV to smooth out the optical response. The \textsc{yambo} code  was used to perform all quasiparticle calculations \cite{sangalli2019}.

To quantify the percentage of the electronic distribution in the layer where the hole is located, we performed a numerical integration of the real-space electron density over the coordinates perpendicular to the $z$-direction, yielding the projected electron density. This projected distribution is then integrated up to a threshold set halfway between the top and bottom layers, and normalized by the value resulting from the integration of the projected density from 0 to $L_z$, where $L_z$ denotes the lattice parameter along the $z$-direction. This approach is similar to the method used in Reference \cite{reho2024excitonic} to evaluate electronic distribution in TMDC heterostructures.

The exciton radiative lifetime at a given temperature is obtained following the model proposed by Spataru \emph{et al}., which has been successfully used to predict lifetimes in semiconducting carbon nanotubes \cite{spataru2005theory}. In this regard, the finite temperature ($T$) radiative lifetime $\tau_{S}(T)$ of the $S$-\emph{th} exciton is defined as the inverse of the radiative rate,
\begin{equation}
	\tau^{-1}_{S}(T)=\langle \gamma_{S}(T)\rangle= \gamma(0) \frac{4}{3}\frac{\Omega_{S}}{\sqrt{2\pi M_{S}c^{2}k_{B}T}},
	\label{eq4}
\end{equation}
where $M_S$ is the effective exciton mass, computed as the sum of the electron and hole masses, while the radiative decay rate for $Q$=0 is given by $\gamma(0)= 2\pi e^{2} \Omega_{S}^{2}\mu_{S}^{2}/(\hbar^{3}c^{2}a_{0})$. Here $a_0$ is the unit cell size along the ribbon axis, and  $\mu_S=\sum_{vck}A_{vck}^{S}d_{cvk}$ represents the excitonic transition dipole matrix elements, being $d_{vck}$ the electronic dipoles, and $\Omega_S$  represents the exciton energy computed at zero momentun transfer (Q=0). 
\begin{figure}[t!]
	\centering
	\includegraphics[width=0.75\linewidth]{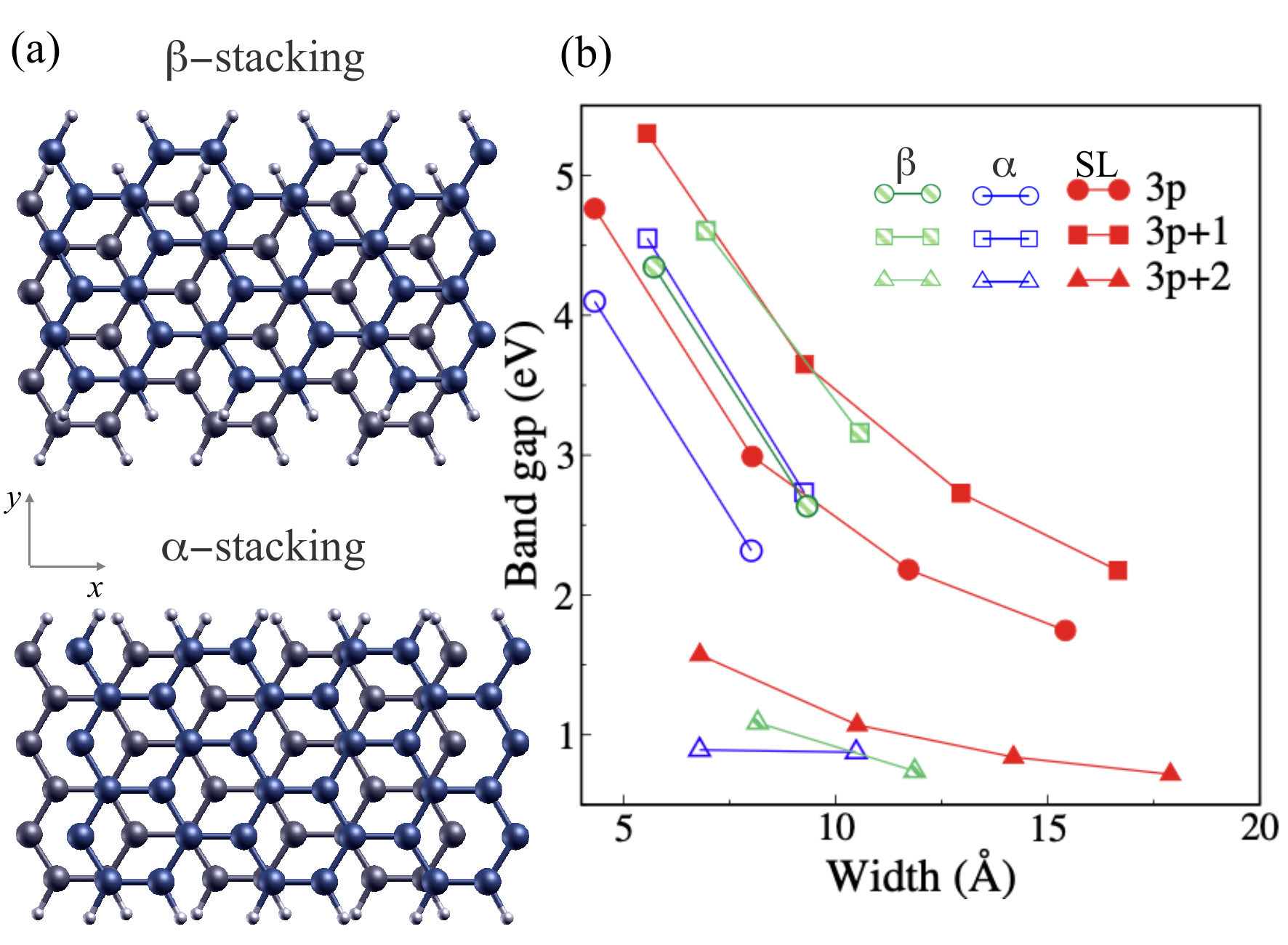}
	\caption{(a) Schematic representation of bilayer 6-AGNRs forming the $\beta$-stacking (top panel) and $\alpha$-stacking (bottom panel). (b) Quasiparticle band gap versus the ribbon width for single-layer (SL) and bilayer AGNRs of different families. The bilayer systems with $\alpha$-stacking possess widths that resamble the single layer ones while the $\beta$-stacking are slightly wider due to the relative shift of the top layer with respect to the bottom one. The dot, square, and triangular symbols indicate the AGNR family to which each ribbon belongs: $3p$, $3p+1$, and $3p+2$, respectively. The filled, unfilled, and diagonally striped versions of these symbols represent the type of ribbon: single-layer, $\alpha$-stacked, and $\beta$-stacked, respectively.}
	\label{fig1}
\end{figure}
\section{Results and Discussion}
\subsection{Homobilayers}
We initially consider two types of bilayer AGNR systems with a Bernal stacking and ideal hydrogen-passivated armchair edges, as depicted in Figure \ref{fig1}\textcolor{blue}{a}. These configurations, commonly referred to as $\beta$- and $\alpha$-stackings, differ in how the top layer is shifted relative to the bottom layer \cite{sahu2008energy}.

We begin our discussion by addressing the effects of electron-electron interactions on the electronic properties, and analyzing the impact of different stackings on the QP band gap of homobilayers.  As shown in Figure \ref{fig1}\textcolor{blue}{b}, the QP band gap of bilayer AGNRs with $\beta$-alignment is slightly higher than that of $\alpha$-alignment but lower than their corresponding single-layer. This behavior is expected, as the dielectric screening introduced by the adjacent layer in bilayer structures reduces the band gap. For the narrowest bilayer systems this leads to a reduction of at least 10\% compared to the single-layer case.  Overall,  the bilayer structures display QP band gap renormalization that is at least 2.5 times larger than the GGA values. This is a signature of the key role played by the Coulomb interactions in these systems with reduced dimensionality (see Table S1 in the Supporting Information).

Given that single-layer 6-AGNR is highly appealing for photovoltaic applications ---its optical band gap is optimal for photon-to-electrical current generation \cite{ruhle2016tabulated}---we use it as a building block to construct the bilayer systems in this work, and we label them 6$\alpha$- and 6$\beta$-stacking.
The lattice constant variation across different stackings is smaller than 0.03\%, indicating that stacking order has a negligible effect on the in-plane lattice constant, which is approximately 4.31 \AA{}. Moreover, we predict interlayer distances of 3.32 \AA{} and 3.30 \AA{} for the $\alpha$- and $\beta$-alignments, respectively. Both structures are energetically favorable, with the 6$\alpha$-stacking being slightly more stable by 0.26 meV/atom. These energy differences are consistent across other families of bilayer GNRs (see Table S1).  

\begin{figure*}[t!]
	\centering
	\includegraphics[width=1.0\linewidth]{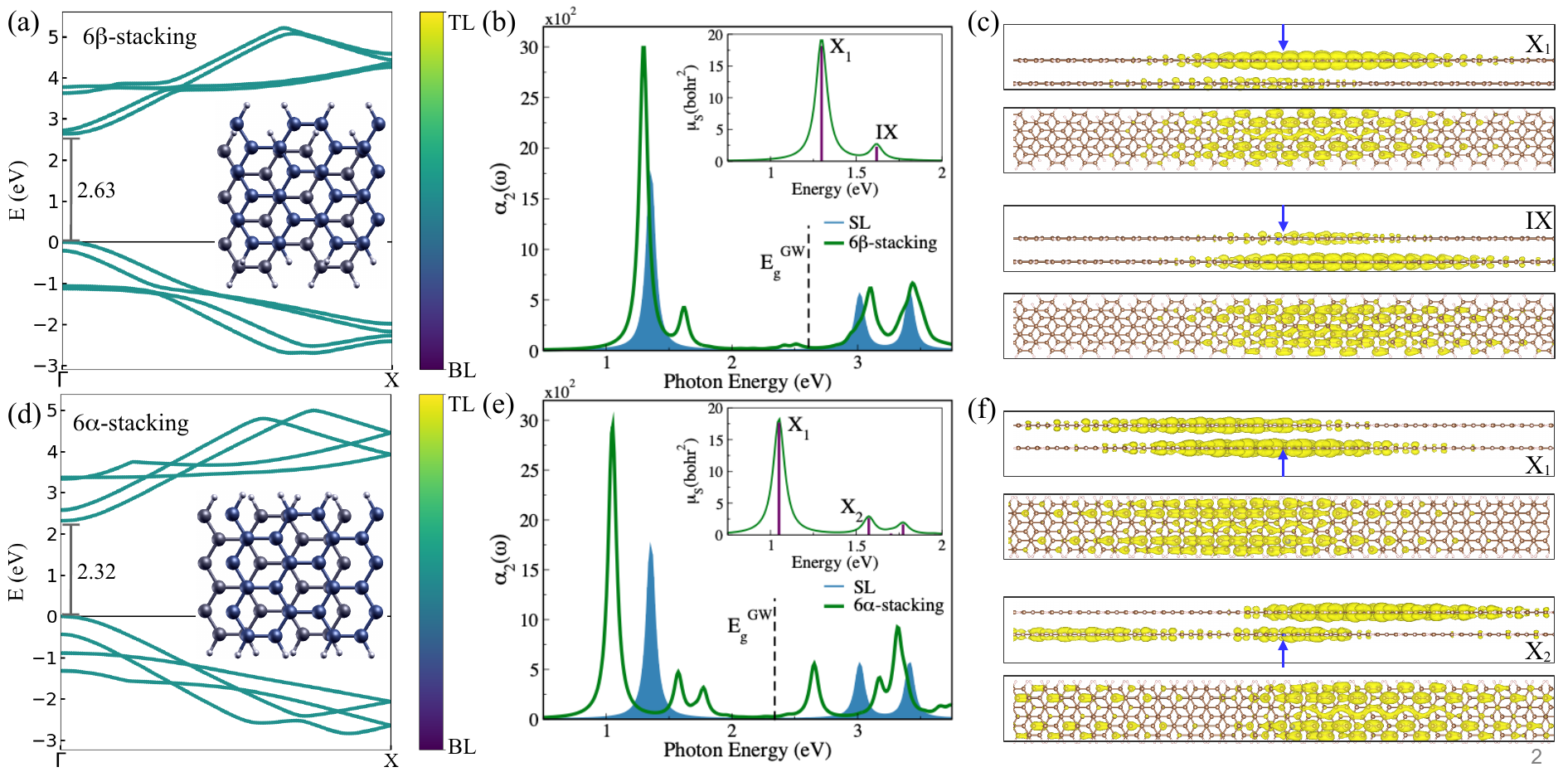}
	\caption{Layer-projected QP band structure for (a) 6$\beta$- and (d) 6$\alpha$-stacking structures. TL and BL indicates the projection on the top and bottom layer, respectively. The absorption spectra including electron-hole interactions for (b) 6$\beta$, and (e) 6$\alpha$ configurations. The light blue filled curve represents the optical absorption of the single-layer 6-AGNR. The insets present a zoon-in view of the lowest-energy normalized excitonic transition dipoles $d_{vck}$. The side and top view of the real-space electron distribution around the hole position (indicated by the blue arrow), for the two-lowest energy excitonic states, are  presented in (c) for 6$\beta$-stacking, and (f) for 6$\alpha$-stacking.}
	\label{fig2}
\end{figure*}

The quasiparticle layer-projected band structures are shown in Fig. \ref{fig2}\textcolor{blue}{a,d}. Regardless of the stacking, strong hybridization can be observed, primarily attributed to the dominant C $p$-states. In addition to the difference in the QP electronic band gap—0.31 eV higher for 6$\beta$-stacking—the main differences in the electronic bands lie in the enhanced relative splitting of the bands within both the lowest-energy conduction bands and the highest-energy valence bands for 6$\alpha$-stacking.

Solving the Bethe-Salpeter equation, allows us to compute the imaginary part of the polarizability per unit length ($\alpha_2$), which provides access to the absorption spectra associated with direct band transitions.
As shown in Fig. \ref{fig2}\textcolor{blue}{b}, the absorption onset for 6$\beta$-stacking displays a bright, lowest-energy exciton at X$_{1}$$\sim$ 1.30 eV, primarily arising from transitions between the valence band maximum (VBM) and the conduction band minimum (CBM) around $\Gamma$. Additionally, the excitonic spectra resemble that of an isolated single-layer 6-AGNR, with only slight deviations and the emergence of an excitonic state at IX $\sim$1.62 eV. This new state predominantly originates from transitions between the second-highest valence band and the second-lowest conduction band.
In contrast, the absorption profile of the 6$\alpha$-stacking, shown in Fig. \ref{fig2}\textcolor{blue}{e}, differs significantly from the absorption spectra of the single-layer 6-AGNR.  Although the nature of optical excitations arises from a complex interplay between interlayer hybridization and structural constraints, the differences observed in the optical spectra can largely be attributed to the edge configurations. For example, in the $\beta$ phase, the outer edge atoms are generally more spatially separated from those in the adjacent layer compared to the $\alpha$ phase. As a result, the electronic charge density at the band edge in the top and bottom layers of the $\beta$ phase distributes in a manner that minimizes interlayer electronic interactions (see Fig. S1 in the Supplementary Information), indicating weaker interlayer coupling. This leads to reduced band repulsion, thereby preserving key features of the 6-AGNR monolayer spectrum. In contrast, the stronger interlayer coupling in the $\alpha$ phase enhances band repulsion, introducing additional features in the optical response.
For the 6$\alpha$-stacking, the lowest-energy excitations are X$_1$$\sim$ 1.05 eV and X$_2$$\sim$ 1.57 eV.  The exciton binding energies of these states are 1.34 eV and 1.30 eV, respectively. Similarly, for 6$\beta$-stacking, a binding energy of 1.33 eV is estimated for X$_1$.

To analyze the real-space exciton localization, we plot the excitonic wave function. For 6$\beta$-stacking, as shown in Fig. \ref{fig2}\textcolor{blue}{c}, X$_1$ localizes approximately 81\% of the electron in the same layer where the hole is located, indicating its intralayer character. On the other hand, the IX excitonic state concentrates 78\% of electrons in the layer opposite from the, suggesting its interlayer character. It presents a non vanishing oscillator strength of $\sim$13\% compared to exciton X$_1$, a feature that may facilitate its application for optoelectronic applications. We have verified that, the observation of interlayer excitons in $\beta$-stacking remains a consistent feature also in bilayers formed by 3-, 4-, and 5-AGNRs (see Supporting Information).

In Fig. \ref{fig2}\textcolor{blue}{f}, we present the exciton wavefunction for  6$\alpha$-stacking. The first two excitonic states exhibit electron distributions that spread across both the bottom and top layers, suggesting a hybridized excitonic nature. Note that the node feature in the bottom layer for X$_2$ is a direct consequence of the presence of pockets in the $k$-space modulus squared of the exciton wavefunction (see Fig. S1 in the Supporting Information). It should be mentioned that we tested multiple hole positions in the most possibly symmetric points of the lattice with the highest electron density in the valence band orbitals that participate in the excitation, all of which resulted in the same asymmetric distribution. Furthermore, we verified that none of the studied systems exhibit degenerate states in their lowest-energy excitonic transitions. This rules out the possibility that the observed asymmetry in the real-space exciton wavefunction is related to degenerate states, as previously demonstrated by Wirtz et al. for h-BN \cite{wirtz2008comment}.
The overall asymmetry and nodal electron distribution with respect to the hole position have also been reported in other single-layer and bilayer graphene nanoribbons (GNRs) featuring either irregular edge geometries or by stacking additional layers on top of single-layer GNRs \cite{zhu2010excitons,wang2012quasiparticle,lou2017quasiparticle,ge2022first}. These findings suggest that the edge configuration plays a key role in generating asymmetric electron densities—an effect that is not commonly observed in other two-dimensional systems.

It is worth noting that the excitons observed in homobilayer AGNRs are in fact hybridized excitons, formed through the coupling of intralayer excitons with optically dark charge-transfer excitons—i.e., excitations in which the electron and hole reside on different layers. Overall, this mechanism is similar to that observed in non-centrosymmetric bilayer MoS$_2$ \cite{deilmann2018interlayer}.
To unambiguously understand the exciton formation mechanism in homobilayer GNRs, one can include spin–orbit coupling (SOC) effects and analyze the resulting optical excitations. Considering that such excitations must obey spin-selection rules—and given that the valence band maximum (VBM) comprises two degenerate bands, one associated with the bottom layer and the other with the top layer, each with opposite spin orientations (a similar configuration exists for the conduction band minimum)—the nature of the optical transitions can be systematically determined (see Fig. S9 of the SI). 
More importantly, because the optical spectra and electron densities are unchanged by the inclusion of SOC, we emphasize that the identification of hybridized excitons—whether through spin-selection rules or through real-space exciton distributions—leads to the same conclusions.

The exciton radiative lifetimes, at room temperature, for 6$\beta$-stacking, $\tau_{S}$ of the intralayer exciton is 140 ps, which is slightly smaller than the one of the first exciton state in the single-layer 6-AGNR (170 ps). These results are in good agreement with previous experimental studies, which suggest exciton lifetimes exceeding 100 ps for sub-2-nm graphene nanoribbons \cite{tries2020experimental}. More importantly, for the interlayer exciton (IX), we estimate significantly longer lifetimes of 1.66 ns, approximately 12 times greater than that of the intralayer exciton.
At the same time, for $\alpha$-stacking, the lifetimes of the first and second hybridized exciton states are 150 ps and 2.05 ns, respectively.
\begin{figure*}[t!]
	\centering
	\includegraphics[width=1.0\linewidth]{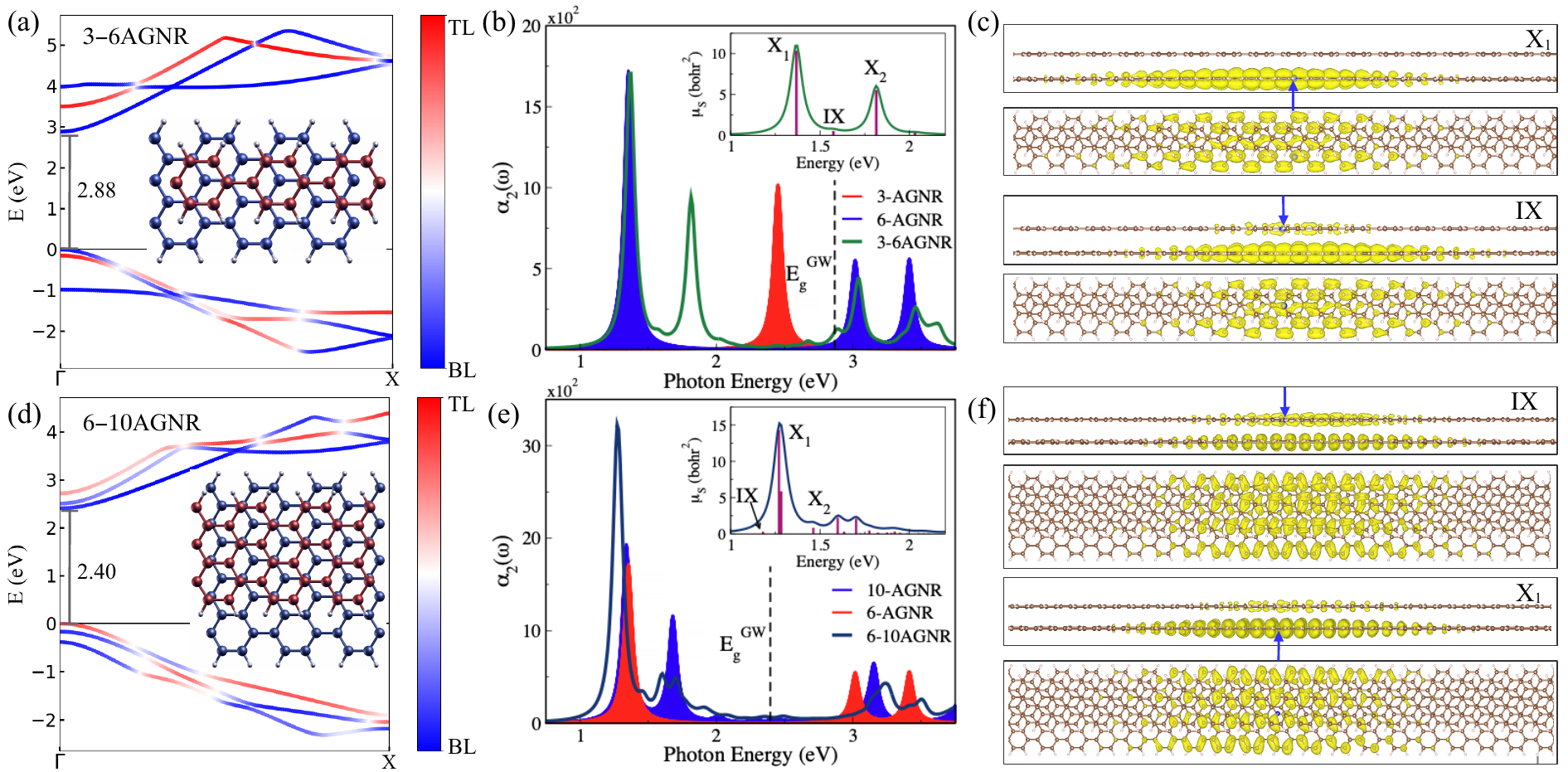}
	\caption{Layer-projected quasiparticle band structure for (a) 3-6AGNR and (d) 6-10AGNR stacking bilayer structures. The first (second) index corresponds to the width of the top (bottom) layer. TL and BL indicates the projection on the top and bottom layer, respectively. The absorption spectra including electron-hole interactions for (b) 3-6AGNR, and (e) 6-10AGNR configurations. The insets present the lowest-energy normalized excitonic transition dipoles $d_{vck}$. Side and top view of the real space electron distribution around the hole position (indicated by the blue arrow) for the two-lowest energy excitonic states corresponding to the (c) 3-6AGNR, and (f) 6-10AGNR bilayer systems.}
	\label{fig3}
\end{figure*}
\subsection{Heterobilayers}
We now consider systems constructed by stacking single-layer AGNRs of different widths (3-, 6-, and 10-AGNRs). It should be noted that the selection of heterobilayer systems was based on prior knowledge of the quasiparticle (QP) band gaps of the individual monolayers \cite{villegas2024screened}. This information enabled us to select monolayers whose QP band edges differ by only a few hundred meV, providing a first-order approximation of the expected interface type (type-II being the most relevant).
For example, the QP band gaps of monolayer 6-AGNR and 10-AGNR are 2.99 eV and 2.73 eV, respectively. The small difference between these values (~0.25 eV) facilitates the formation of a heterobilayer with type-II band alignment. In contrast, the QP gap difference between 3-AGNR and 6-AGNR is significantly larger (1.77 eV), leading to a heterobilayer with type-I alignment.
While other combinations with favorable band gaps and small QP differences could, in principle, be explored, the large number of atoms involved—combined with our stringent convergence requirements for the $k$-point mesh, number of bands, and energy cutoff—made such calculations computationally prohibitive for us.
Henceforth, we focus on the 3-6AGNR and 6-10AGNR, which are schematically represented in the insets of Figure \ref{fig3}\textcolor{blue}{a,d}, respectively. The fully relaxed systems exhibit features similar to the homobilayer $\beta$-alignment along one edge, i.e, the bottom and top layer are shifted one with respect the other. It is worth mentioning that regardless of the initial placement of the top layer, the systems consistently adopt a $\beta$-like stacking after relaxation. 

In Fig. \ref{fig3}\textcolor{blue}{a,d} we present the quasiparticle layer-projected band structures. It is evident that the VBM and CBM around the $\Gamma$ point show no signs of hybridization between layers. For the 3-6AGNR system, this enables the localization of both the VBM and CBM within the 6-AGNR layer (bottom layer). Similarly, the second-lowest conduction band and the second-highest valence band are localized in the 3-AGNR layer (top-layer). Consequently, the band alignment of this heterobilayer system corresponds to type-I.
In contrast, for the 6-10AGNR heterobilayer, the VBM is localized in the 6-AGNR layer (top-layer), while the CBM is localized in the 10-AGNR layer (bottom-layer), resulting in a type-II band alignment. 
The QP band gaps for these heterobilayer systems vary from 2.40 eV to 2.88 eV, with the narrower system exhibiting a larger band gap, as expected.

Regarding the excitonic effects, the lowest-energy exciton spectra for the 3-6AGNR heterobilayer, shown in Fig. \ref{fig3}\textcolor{blue}{b}, exhibits a bright exciton state, X$_1$, at 1.37 eV. This state originates from transitions between the VBM and the CBM around the $\Gamma$ point. Note that the excitonic transition dipole moment of this state is reduced by half compared to those in the 6$\alpha$- and 6$\beta$-stacking configurations (see  inset of Fig. \ref{fig3}\textcolor{blue}{c}). By plotting the real-space exciton wavefunction, as depicted in Fig. \ref{fig3}\textcolor{blue}{c} (top panel), we confirm that it corresponds to an intralayer state, with 98\% of the electron distribution concentrated around the hole position.
Additionally, we identify the presence of an interlayer exciton, IX, at 1.57 eV (see the inset of Fig. \ref{fig3}\textcolor{blue}{c}), which arises from transitions between the second valence band maximum and the CBM near the $\Gamma$ point. The exciton wavefunction further verifies its interlayer character, with 92\% of the electron distribution localized in the layer opposite to the hole position.
It is worth noting that the optical response of the heterobilayer largely retains the excitonic features characteristic of the isolated single-layer 6-AGNR. This behavior is expected, given the narrower width of the top layer (3-AGNR), which induces only weak screening effects on the bottom layer (6-AGNR), thus preventing significant changes in its QP properties. Quantitatively, this can be seen as a cancellation effect between the QP band gap correction and the exciton binding energy for the lowest excitonic state (see Table~\ref{tab1}).  In contrast, the excitonic state of the heterobilayer associated with the top layer (3-AGNR), at $\sim$1.8 eV,  exhibits a significant redshift relative to the lowest-energy exciton of the isolated 3-AGNR monolayer. These results suggest that the enhanced screening in the bilayer system primarily affects the narrower of the two layers.

The absorption spectrum of the 6-10AGNR heterobilayer, shown in Fig. \ref{fig3}\textcolor{blue}{e}, presents a bright exciton at 1.29 eV, originating from transitions between the second-highest valence band to the lowest conduction band. Its intralayer character is confirmed by plotting the real-space exciton wavefunction (see bottom panels in Fig. \ref{fig3}\textcolor{blue}{f}), where 92\% of the electrons are distributed in the same layer as the hole.
A closer inspection of the spectra, however, reveals that the lowest-energy excitonic state, IX = 1.18 eV, corresponds to an interlayer exciton ---80\% of the electronic distribution in the opposite layer from the hole--- with an oscillator strength that is $\sim$0.25\% of exciton X$_1$ as depicted in the inset of Fig. \ref{fig3}\textcolor{blue}{e}.

We then estimate the exciton radiative lifetimes for both heterobilayer systems. At room temperature, considering the 3-6AGNR system, we calculated exciton lifetimes of 160 ps and 258 ns for the intralayer and interlayer excitons, respectively. For the 6-10AGNR system, we predicted that the lowest-energy intralayer exciton possesses a radiative lifetime of 180 ps, while the lifetime of the interlayer exciton (IX) extends to thousands of nanoseconds (9.39 $\mu$s). The extremely long interlayer radiative lifetimes related to the 6-10AGNRs  may be ascribed due to its smaller oscillator strengths and large exciton mass, when compared with the homobilayer and 3-6AGNR system.
\begin{table*}[t!]
	\caption{ Predicted band gap at the G$_0$W$_0$ level, exciton binding energy (E$_b$), singlet-triplet splitting energy ($\Delta^{ST}$), and intrinsic ($\tau_{0}=\gamma_{0}^{-1}$) and room temperature ($\tau_{s}^{\text{RT}}$) radiative lifetimes for different single-layer and bilayer AGNRs. The exciton binding energies correspond, in all cases, to the lowest-energy excitonic state. The exciton mass M$_S$ is also included.}
	\label{tab1}
	\centering
			\begin{ruledtabular}
			\renewcommand{\arraystretch}{1.1}
				\resizebox{\textwidth}{!}{\begin{tabular}{@{}llllllllll@{}}
						&  &  &  & \multicolumn{3}{c}{Intralayer}  & \multicolumn{3}{c}{Interlayer}  \\
						\cline{5-7}\cline{8-10}
						&
						E$_{g}^{\text{GW}}$ (eV) &
						E$_{b}$ (eV) &
						$\Delta^{ST}$ (eV) &
						$M_{\text{S}}$  (m$_0$)&
						$\tau_{0}$ (ps) &
						$\tau_{s}^{\text{RT}}$ (ns) &
						$M_{\text{S}}$ (m$_0$) &
						$\tau_{0}$ &
						$\tau_{s}^{\text{RT}}$ (ns) \\\hline 
						3-AGNR          & 4.76 & 2.33 & 0.67 & 0.292 & 1.83 & 90 &  -  & - & - \\
						6-AGNR           & 2.99 & 1.625 & 0.29 & 0.233 & 2.23   & 170 &  -  & - &-\\
						10-AGNR          & 2.73 & 1.43 & 0.21 & 0.462 & 1.85   & 200 &  -  & - &-\\
						6$\alpha$-stacking  & 2.32 & 1.27 & 0.20 & 0.347 & 1.27   & 150 &  -  & -& -\\
						6$\beta$-stacking   & 2.63 & 1.33 & 0.23 & 1.144 & 0.79   & 140 & 0.152 & 0.032  & 1.66\\
						3-6AGNR          & 2.88 & 1.51 & 0.26 & 0.228& 2.18  & 160 & 0.284 & 3.52  & 258\\
						6-10AGNR         & 2.40 & 1.22 & 0.14 & 0.398 & 4.34  & 180 & 0.377& 83.7  & 9.4$\times 10^3$ 
					\end{tabular}
				}
					\end{ruledtabular}
		\end{table*}
		
		Finally, we calculate the singlet–triplet splitting ($\Delta^{ST}$) of the studied systems, defined as the energy difference between the lowest singlet and lowest triplet state. For the bilayer systems, $\Delta^{ST}$ range from 0.14 to 0.26 eV, while the narrowest (widest) single-layer present a value of 0.67 eV (0.21 eV). The value for narrowest single-layer (3-AGNR) is comparable to that of organic polymeric systems \cite{wilson2000triplet}, while the value for the widest ribbon is similar to that of flat \ce{MoS2} nanoribbons \cite{tang2022tunable}. 
		In principle, the relatively large values of $\Delta^{ST}$ found in the bilayer systems help suppress the intersystem crossing (ISC) process, enabling the retention of high singlet exciton populations and, consequently, enhancing electroluminescence efficiency \cite{shuai2000singlet,chen2006singlet}. The values for the singlet–triplet splitting as well as the exciton radiative lifetimes are summarized in Table \ref{tab1}.
		
		\subsection{Discussion}
		Overall, our findings highlight the intriguing properties of bilayer AGNRs, which make them promising semiconductor systems for designing novel optoelectronic devices.
		First, currently quasi‐free‐standing bilayer GNRs with AB‐stacking have been synthesized by combining epitaxial growth of GNRs with air annealing processes \cite{oliveira2015synthesis}. Indeed, this method enables modulation of GNR widths by varying either the growth temperature or the growth time, paving the way for large‐scale fabrication of bilayer GNRs with a wide variety of homobilayer and heterobilayer configurations. In these structures, type-I and type-II vertical heterojunctions with unique excitonic characteristics can be explored, as demonstrated here.
		Although type-II heterojunctions are diserable for practical optoelectronic applications, we mention that a recent study conducted in TMDC-based heterobilayer systems has demonstrated that type-I band alignments can be electrically engineer to form type-II ones  \cite{kistner2024electric}. 
		
		Our first-principles simulations suggest that the investigated bilayer AGNRs exhibit strong optical responses that cover the infrared to visible range of the electromagnetic spectrum, Exciton binding energies reaching hundreds of meV, making them robust against thermal fluctuations. 
		We argue that such strong electron-electron/electron-hole correlations may also play a crucial role in other Q1D graphene-based systems such electronic waveguides \cite{rickhaus2013ballistic,zhang2009guided,villegas2010comment,villegas2012controlling}
		
		Second, while the exciton lifetime for the 1.7 nm wide single-layer graphene nanoribbons has been reported to exceed 100 ps \cite{tries2020experimental}, our predicted radiative exciton lifetime for a 1.3 nm wide single-layer graphene nanoribbon (10-AGNR) can be considered a reasonable estimation of radiative rates, validating the reliability of our theoretical results. Based on this, we argue that long-lived interlayer excitons, with radiative lifetimes ranging from 1 ns to 9.4 $\mu$s, can be observed in homobilayer and heterobilayer carbon-based systems. We should highlight that our simulations assume defect-free samples. As such, the computed excitation energies and radiative lifetimes may differ quantitatively to experimental conditions due to factors like substrate choice or the presence of defects. For low defect concentrations ($<$0.7\%), band-edge states — key to the lowest-energy excitations — are only slightly affected \cite{rojas2019ab}, suggesting that their main features undergo only minor modifications. At higher concentrations, these states shift significantly, altering conduction channels and possibly the excitonic behavior. Since defects also introduce non-radiative recombination paths, it is likely that shorter exciton lifetimes would be experimentally reported. Hence, our predicted lifetimes should be seen as upper bounds, serving as benchmarks for future experimental studies.
		
		Note also that our study has considered only freestanding systems, although real semiconductor nanoribbons are typically surrounded by an underlying substrate that creates a polarized environment. While the effect of substrates has not been considered in this study, it is expected that surrounding dielectric materials will considerably reduce both the QP band gap, the optical excitation energy, and the binding energies \cite{raja2017coulomb,garcia2011renormalization}. In this regard, we mention that our recently proposed effective model for 1D semiconductors \cite{villegas2024screened} can be extended to consider substrate effects following the procedure proposed by Riis-Jensen et al. for 2D systems \cite{riis2020anomalous}.
			
		It is worth mentioning that future research on wider carbon-based heterobilayer ribbons—which require the simulation of unit cells containing a large number of atoms—may benefit from the time-dependent density functional theory (TDDFT) approach developed by Wing et al \cite{wing2019comparing}. This method employs screened range-separated hybrid (SRSH) functionals, addressing the underestimation of the ground-state band gap while satisfying the correct long-range behavior of the exchange-correlation kernel. As a result, it provides optical responses that closely match those obtained from GW-BSE calculations, but at a significantly reduced computational cost.
		
		For bilayer systems composed of mixed AGNR-, ZGNR-, and chevron-like monolayers that exhibit non-hybridized excitons, it would be valuable to conduct a comprehensive analysis of the symmetry properties of both the wave functions and the optical selection rules. Such an analysis could follow the approaches used in studies of twisted bilayer graphene quantum dots \cite{wang2022polarization} or other 2D heterostructures \cite{zhong2023exciton}. 
		
		While we recognize the experimental challenges that must be overcome to realize bilayer GNR systems—including precise control over edge shape and lateral width, as well as compatibility with printed electronics—our findings pave the way for the experimental realization of next-generation optoelectronic devices based on bilayer graphene nanoribbons.
		
		\section{Conclusions}
		We investigated the quasiparticle electronic and excitonic properties of carbon-based homobilayer and heterobilayer systems, as well as the exciton radiative rates for different stacking configurations. The results indicate that homobilayer AGNRS with $\beta$-stacking are prone to exhibiting intralayer and interlayer excitons. In contrast, mixed hybridized excitons, characterized by an electron distribution across both the bottom and top layers, were observed in $\alpha$-stack systems.
		For heterobilayer systems, the fully relaxed structures adopted $\beta$-like stacking, forming both type-I and type-II band alignments. This facilitates the formation of pure intralayer and interlayer excitons, with the latter exhibiting long-lived room-temperature radiative lifetimes of 258 ns and 9.4 $\mu$s for the 3-6AGNR and 6-10AGNR systems, respectively.
		These findings highlight the possibility of engineering interlayer exciton lifetimes and their binding energies by controlling the stacking sequence of carbon-based bilayer systems, which are critical properties for the development of novel optoelectronic and excitonic devices.

		\section{Acknowledgements}
		The authors acknowledge financial support from the Brazilian agencies CAPES(001), CNPq, FAPERJ, and FAPES (TO-1043/2022). ARR acknowledges support from FAPESP (Grant No. 2017/02317-2 and 2021/14335-0).  This research was supported by resources supplied by CENAPAD-SP,  and the Center for Scientific Computing (NCC/GridUNESP) of the UNESP.
\bibliographystyle{apsrev}

\end{document}